\documentclass[twocolumn,aip,reprint]{revtex4-1}

\usepackage{dcolumn}
\usepackage{amsmath}
\usepackage{graphicx}
\usepackage{latexsym}  
\usepackage{amsfonts}  
\usepackage{amssymb}
\usepackage{color}  
\usepackage{bbm}
\usepackage{hyperref}
\definecolor{darkblue}{rgb}{0.0,0.0,0.3}
\hypersetup{colorlinks,breaklinks,
            linkcolor=blue,urlcolor=blue,
            anchorcolor=blue,citecolor=blue}
\DeclareGraphicsExtensions{.eps,.pdf,.png,.gif,.jpg}

\begin{document}

\newcommand{\be}{\begin{equation}}
\newcommand{\ee}{\end{equation}}
\newcommand{\bea}{\begin{eqnarray}}
\newcommand{\eea}{\end{eqnarray}}
\newcommand{\ra}{\rightarrow}
\newcommand{\callH}{\mathcal{H}}
\newcommand{\callF}{\mathcal{F}}
\newcommand{\ket}{\rangle}
\newcommand{\bra}{\langle}
\newcommand{\ha}{\hat{a}}
\newcommand{\hu}{\hat{u}}
\newcommand{\hw}{\hat{w}}
\newcommand{\pp}{\textrm{p}}
\newcommand{\had}{\hat{a}^{\dagger}}
\newcommand{\hH}{\hat{H}}
\newcommand{\hV}{\hat{V}}
\newcommand{\hA}{\hat{A}}
\newcommand{\hX}{\hat{X}}
\newcommand{\hhA}{\hat{\mathcal{A}}}
\newcommand{\hU}{\hat{U}}
\newcommand{\hT}{\hat{T}}
\newcommand{\hW}{\hat{W}}
\newcommand{\hv}{\hat{v}}
\newcommand{\hG}{\hat{\mathcal{G}}}
\newcommand{\hh}{\hat{h}}
\newcommand{\hD}{\hat{\Delta}}
\newcommand{\hn}{\hat{n}}
\newcommand{\hpsi}{\hat{\psi}}
\newcommand{\hpsid}{\hat{\psi}^{\dagger}}
\newcommand{\hGamma}{\hat{\Gamma}}
\newcommand{\bh}{\mathbf{h}}
\newcommand{\bj}{\mathbf{j}}
\newcommand{\bG}{\mathbf{G}}
\newcommand{\bA}{\mathbf{A}}
\newcommand{\br}{\mathbf{r}}
\newcommand{\bz}{\mathbf{z}}
\newcommand{\bR}{\mathbf{R}}
\newcommand{\bx}{\mathbf{x}}
\newcommand{\by}{\mathbf{y}}
\newcommand{\bu}{\mathbf{u}}
\newcommand{\ba}{\mathbf{a}}
\newcommand{\bs}{\mathbf{s}}
\newcommand{\bk}{\mathbf{k}}
\newcommand{\bq}{\mathbf{q}}
\newcommand{\bp}{\mathbf{p}}
\newcommand{\bS}{\mathbf{\Sigma}}
\newcommand{\half}{\frac{1}{2}}
\newcommand{\prt}{\partial}
\newcommand{\ua}{\uparrow}
\newcommand{\da}{\downarrow}
\newcommand{\dr}{d^3}
\newcommand{\xc}{\textrm{xc}}
\newcommand{\x}{\textrm{x}}
\newcommand{\cc}{\textrm{c}}
\newcommand{\s}{\sigma}
\newcommand{\m}{{m}}
\newcommand{\drm}{d^{3m}}
\newcommand{\co}{\textrm{c}}
\newcommand{\Ha}{\textrm{H}}
\newcommand{\F}{\textrm{F}}
\newcommand{\Hxc}{\textrm{Hxc}}
\newcommand{\HK}{\textrm{HK}}
\newcommand{\LDA}{\textrm{LDA}}
\newcommand{\GGA}{\textrm{GGA}}
\newcommand{\GEA}{\textrm{GEA}}
\newcommand{\mom}{\hat{\mathbf{p}}}
\newcommand{\kf}{k_{\textrm{F}}}
\newcommand{\ef}{\epsilon_{\textrm{F}}}
\newcommand{\tpth}{(2 \pi)^3}
\newcommand{\bru}{\underline{\br}_m}
\newcommand{\bqu}{\underline{\bq}_m}
\newcommand{\dru}{d \bru}
\newcommand{\drd}{d^D}
\newcommand{\tpd}{(2 \pi)^D}
\newcommand{\tpt}{(2 \pi)^2}
\newcommand{\mf}{\mathcal{F}}
\newcommand{\llangle}{\langle \! \langle}
\newcommand{\rrangle}{\rangle \! \rangle}
\newcommand{\w}{\omega}

\title{Ultra-nonlocality in density functional theory for photo-emission spectroscopy}

\author{A.-M. Uimonen}
\affiliation{Department of Physics, Nanoscience Center, University of Jyv\"askyl\"a, Survontie 9, 40014 Jyv\"askyl\"a, Finland}
\author{G. Stefanucci}
\affiliation{Dipartimento di Fisica, Universit{\'a} di Roma Tor Vergata, Via della Ricerca Scientifica, 00133 Rome, Italy}
\affiliation{INFN, Laboratori Nazionali di Frascati, Via E. Fermi 40, 00044, Frascati, Italy}
\affiliation{European Theoretical Spectroscopy Facility (ETSF)}
\author{R. van Leeuwen}
\affiliation{Department of Physics, Nanoscience Center, University of Jyv\"askyl\"a, Survontie 9, 40014 Jyv\"askyl\"a, Finland}
\affiliation{European Theoretical Spectroscopy Facility (ETSF)}
\date{\today}

\begin{abstract}
We derive an exact expression for the photocurrent of photo-emission spectroscopy using 
time-dependent current density functional theory (TDCDFT). This expression is given as an 
integral over the Kohn-Sham spectral function renormalized by effective potentials that depend 
on the exchange-correlation kernel of current density functional theory.
We analyze in detail the physical content of this expression by making a connection between 
the density-functional expression and the diagrammatic expansion of the photocurrent within 
many-body perturbation theory. We further demonstrate that the density functional expression 
does not provide us with information on the kinetic energy distribution
of the photo-electrons. Such information can, in principle, be obtained from TDCDFT by exactly 
modeling the experiment in which the photocurrent is split into energy contributions by means 
of an external electromagnetic field outside the sample, as is done in standard detectors.
We find, however, that this procedure produces very nonlocal correlations between the exchange-correlation fields
in the sample and the detector.

\end{abstract}
\pacs{31.15.E-, 31.15.ee, 31.15.eg }
\maketitle


\section{Introduction}

The photo-electric effect, in which electrons are emitted from a material by applying light, 
has played an important conceptual role in quantum mechanics. Already in 1905
Einstein \cite{Einstein} established the famous relation
\be
E_\textrm{K} =  \hbar \omega -\Phi \nonumber
\ee
where $E_\textrm{K}$ is the maximum kinetic energy of the emitted photo-electrons, 
$\hbar \omega$ the energy of the incoming photons and $\Phi$ the work function of the 
material (which is equal to minus the chemical potential provided we use a gauge in 
which the potential is zero at infinity \cite{LangKohn:PRB71}).  Presently photo-emission 
spectroscopy is a well-developed tool for the study of  bandstructures and surface properties 
of materials (for a review see e.g. Ref.~[\onlinecite{Damascelli}])  in which, apart from the kinetic energy, 
also the angular distribution of the photo-electrons is measured. The photo-emission spectrum 
is closely related to the spectral function of the material which can exhibit a wide range
of many-body features such as quasi-particle broadening and plasmon satellites. 
Furthermore there are so-called extrinsic effects describing the energy losses of the 
photo-electron within the material on its way to the surface. The proper treatment of all 
these phenomena requires a many-body description. The underlying theory
has been described in a number of classic references \cite{Almbladh1,Almbladh2,Almbladh3,Hedin1985}. 
Although these many-body approaches can deal with complex many-body processes 
they are computationally expensive. One may therefore wonder whether one could develop
a computationally more efficient approach based on density-functional 
theory \cite{DreizlerGross,Engelbook,Barthreview,Ullrichbook}. 
As the photo-emission process is a time-dependent phenomenon we need a time-dependent version of
density functional theory \cite{Ullrichbook,TDDFTreview,KeldyshTDDFT,Ullrich2,Ruggenthaler}.
Since the outgoing photocurrent density $\bj (\br,t)$ is
a key variable in time-dependent current-density functional theory  (TDCDFT) 
\cite{VignaleKohn,Vignale,Tokatly,Ullrichbook} an approach based on this
formalism appears the most promising. In terms of the time-dependent many-body 
state $| \Psi (t) \rangle$ the current density is given by
\be
\bj (\br,t) = \langle \Psi (t) | \hat{\bj}_\pp (\br ) | \Psi (t) \rangle +  \langle \Psi (t) | \hat{n} (\br ) | \Psi (t) \rangle \bA (\br t)
\ee
where $\bA$ is the applied vector potential, $\hat{n} (\br)$ is the density operator and 
\be
\hat{\bj}_\pp (\br, t) = \frac{1}{2i} \sum_\sigma [ \hpsid (\bx) \nabla \hpsi (\bx) - \nabla \hpsid (\bx) \hpsi (\bx)]  \nonumber
\ee
is the paramagnetic current operator expressed in terms of the field operators 
$\hpsi$ and $\hpsid$, where $\bx = \br, \sigma$ is a space-spin index.
In TDCDFT this current density is calculated instead from a Kohn-Sham 
state $| \Psi_s (t) \rangle$
with a time-evolution determined by a non-interacting Kohn-Sham Hamiltonian $\hH_s (t)$.
This Hamiltonian contains an external Kohn-Sham vector field $\bA_s$ (in a gauge where we absorb the scalar
potentials in a vector potential) which is a functional of the current density.
In this way the photo-emission experiment can be modelled theoretically
by time-propagation of Kohn-Sham orbitals after a suitable approximation for the
Kohn-Sham vector potential $\bA_s$ has been chosen. Indeed some first calculations 
of this kind have been carried out recently \cite{Pehlke}.

There are, however, two issues that remain unresolved. The first issue is the 
question whether TDCDFT allows for the determination of the kinetic energy 
distribution of the photo-electrons. The second issue is what the quality
of the corresponding exchange-correlation kernels must be in order to 
account for many-body features such as plasmon losses.
These are the two issues that we will address in this paper. The paper is divided as follows. 
In section II we briefly review the many-body approach to photo-emission
where we stress the equations that are relevant for the connections to density-functional theory. In Section III
we give a description of TDCDFT in the language of Keldysh theory in order to make connection with
the standard many-body approaches. We further give a discussion of the calculation of the kinetic energy
distribution and the related very long range nonlocalities that are required in TDCDFT to calculate this
property exactly. Finally in Section IV we give our outlook and conclusions.

\section{Many-body theory of photo-emission}
\subsection{The photocurrent}
\label{sec:photocurrent}

Here we will present a short overview of the many-body approach to photo-emission
in which we highlight the aspects relevant to the density functional treatment.
We will follow the approach outlined by Almbladh \cite{Almbladh1}. We assume that the
many-body system is described by a Hamiltonian of the form
\be
\hH (t) = \hH_0 + \hD (t)  \nonumber
\ee
where $\hD$ describes the electromagnetic field applied
for times $t > t_0$ and 
$\hH_0$ is the many-body Hamiltonian of the sample before the field is applied.
The time-evolution of the many-body state is described by the 
time-dependent Schr\"odinger equation
\be
i \prt_t | \Psi (t) \rangle = \hH (t) | \Psi (t) \rangle  \nonumber
\ee 
with initial condition $| \Psi (t_0)\rangle = | \Phi_0 \rangle$.
We take $|\Phi_0 \rangle$ to be the ground state of $\hH_0$,
{\em i.e.}, $\hH_0 | \Phi_0 \rangle = E_0 | \Phi_0 \rangle$ where $E_0$ is
the ground state energy. If we know the state $| \Psi (t) \rangle$ then we
can calculate all observables of interest. In the case of photo-emission 
the observable of interest is the current density outside the sample which
describes the emission of photo-electrons. This amounts to the calculation
of a one-body observable. In non-equilibrium many-body theory \cite{SvL_book} such observables can
be calculated directly from the lesser Green's function defined as
\be
G^< (\bx t, \bx' t') = i \langle \Phi_0 |  \hpsid_H (\bx' t') \hpsi_H (\bx t) | \Phi_0 \rangle  \nonumber
\ee
where $\hA_H(t) = \hU (t_0,t) \hA \hU (t,t_0)$ is the Heisenberg form of the operator $\hA$
with respect to the full Hamiltonian and $\hU$ is the evolution operator of the system which in general is a time-ordered exponential.
The current can then be calculated from
\be
\mathbf{j}_\pp (\br, t) = - \frac{1}{2} \sum_\sigma (\nabla-\nabla') 
G^< (\bx t,\bx' t') |_{\bx=\bx'} .
\label{j_calc}
\ee
Here we concentrate on the paramagnetic part of the current as we will see that the
diamagnetic part only contributes to higher order in the applied field.
Let us see what we get if we expand in powers of the electromagnetic coupling $\hD$. To do this we first expand
the time-dependent many-body state in powers of $\hD$
\be
| \Psi (t) \rangle =\sum_{n=0}^\infty |\Psi^{(n)} (t) \rangle \nonumber
\ee
where $| \Psi^{(n)} \rangle$ is the $n$-th order term. In particular $| \Psi^{(0)} (t) \rangle 
= e^{-i E_0 (t-t_0) } | \Phi_0 \rangle$.  The $k$-th order term in the expectation value for the 
current is then given by
\be
\mathbf{j}^{(k)} (\br,t  ) = \sum_{n+m=k} \langle \Psi^{(n)} (t) | 
\hat{\bj}_\pp (\br) | \Psi^{(m)} (t) \rangle .
\label{currexp}
\ee
If we are interested in the photo-emission current outside the sample then any term with $m=0$ or $n=0$
does not contribute since $| \Psi^{(0)} (t) \rangle$ is localized to the sample in position space and 
vanishes exponentially outside the sample. 
The diamagnetic current  $n (\br t) \bA (\br t )$ is even smaller since
the lowest order contribution not involving $| \Psi^{(0)} (t) \rangle$ is third order in the applied field.
The lowest order non-zero contribution to the photocurrent is therefore
given by
\be
\bj^{(2)}  (\br,t  )  = \langle \Psi^{(1)} (t) | \hat{\bj}_\pp (\br) | \Psi^{(1)} (t) \rangle.
\label{lopc}
\ee 
The other two lowest order terms $\langle \Psi^{(2)} (t) | \hat{\bj}_\pp (\br) | \Psi^{(0)} (t) \rangle$ and
$\langle \Psi^{(0)} (t) | \hat{\bj}_\pp (\br) | \Psi^{(2)} (t) \rangle$  contributing to $\bj^{(2)}$
are zero since we are assuming the photocurrent to be measured far outside the sample.

The calculation of $\bj^{(2)} $ requires the knowledge of the first order change
in the many-body state upon application of the field. This is readily calculated to be
\be
| \Psi^{(1)} (t) \rangle = -i \int_{t_0}^t dt' \, \hU_0 (t,t') \, \hD (t') \, \hU_0 (t',t_0) | \Phi_0 \rangle \nonumber 
\ee
where $\hU_0 (t,t') = e^{-i \hH_0  (t-t')}$ is the time-evolution operator of the unperturbed system.
Using this expression we can write the photocurrent as
\be
\bj^{(2)}  (\br,t  )  = \int_{t_0}^t dt_1 dt_2 \,  \langle \Phi_0 | 
\hD_{H_0} (t_2) \hat{\bj}_{\pp,H_0} (\br  t) \hD_{H_0} (t_1) | \Phi_0 \rangle
\label{photo_curr}
\ee
where the operators are now in the Heisenberg representation with respect to $\hH_0$.
This is the starting expression for all our considerations. In many-body perturbation theory
this expression is expanded in powers of the many-body interactions and can be  
represented as a diagrammatic series. To do this it will be convenient to define the equal-time lesser
Green's function (or equivalently the one-particle density matrix) to second order in the external perturbation
 as
\begin{align}
& G^{(2) <} (\bx t, \bx' t) = \nonumber \\ 
&i \int_{t_0}^t dt_1 dt_2 \,  \langle \Phi_0 | \hD_{H_0} (t_2) \hpsid_{H_0} (\bx't ) \hpsi_{H_0} (\bx t)  \hD_{H_0} (t_1) | \Phi_0 \rangle
\label{GF_exp}
\end{align}
as the Green's function has a well-known expansion in Feynman diagrams. 

\subsection{Diagrammatic expansion}

To expand Eq.~(\ref{GF_exp}) into diagrams it will be convenient to write it as follows
\begin{align}
& G^{(2) <} (\bx t, \bx' t) =  \nonumber \\
&i  \int_{t_0}^t dt_1 dt_2 \,  \langle \Phi_0 | \hD_{+} (t_2) \hpsid (\bx') \hpsi (\bx) \hD_{-} (t_1) | \Phi_0 \rangle
\nonumber
\end{align}
where we defined
\begin{align}
\hD_{-} (t_1) &= \hU_0 (t,t_1) \hD (t_1) \hU_0 (t_1,t_0), \nonumber \\
\hD_{+} (t_2) &= \hU_0 (t_0,t_2) \hD (t_2) \hU_0 (t_2,t)  .\nonumber
\end{align}
The operator $\hD_{-} (t_1)$ can now be expanded in time-ordered
powers of the many-body interaction, whereas $\hD_{+} (t_2)$ can be
expanded in anti-time-ordered powers of the interaction.
Since it will not be our goal to a give an overview of many-body theory we
restrict ourselves to the minimum which is required for understanding the
connections to density functional theory.
 Within 
the language of Keldysh many-body theory \cite{SvL_book} we can say that the operator
$\hD_{-}$ is situated on the forward branch of the Keldysh contour whereas
$\hD_{+}$ is situated on the backward branch. This leads to an expansion
of $G^{(2) <}$ in terms of the non-interacting Greens' functions
\begin{align}
G_{--} (\bx t,\bx' t') &= -i \langle \chi_0 | T [ \hpsi_H (\bx t) \hpsid_H (\bx' t')] | \chi_0 \rangle \nonumber \\
G_{++} (\bx t,\bx' t') &= -i \langle \chi_0 | \tilde{T} [ \hpsi_H (\bx t) \hpsid_H (\bx' t')] | \chi_0 \rangle \nonumber \\
G_{-+} (\bx t,\bx' t') &= \,\, \, i \langle \chi_0 |  \hpsid_H (\bx' t') \hpsi_H (\bx t) | \chi_0 \rangle \nonumber \\
G_{+-} (\bx t,\bx' t') &= -i \langle \chi_0 |  \hpsi_H (\bx t) \hpsid_H (\bx' t') | \chi_0 \rangle \nonumber 
\end{align}
where $T$ is the time-ordering operator, $\tilde{T}$ is a the anti-time-ordering operator
and $| \chi_0 \rangle$ is the ground state of the non-interacting system and the operators
are in the Heisenberg picture with respect to the noninteracting system. 
The Green's functions $G_{-+}$ and $G_{+-}$ are equivalently denoted as $G^<$ and $G^>$.
The vertices in
the diagrams are labeled by $-$ or $+$ depending on whether they lie 
on the forward ($-$)
or backward ($+$) branch of the Keldysh contour.  The bare Coulomb interactions will be denoted
by wiggly lines and since these interactions are instantaneous they will always connect
times on the same branch of the contour.
Often the Green's function lines in the diagrams are dressed by self-energy
insertions such that we can expand in skeleton diagrams ({\em i.e.}, diagrams with self-energy
insertions removed) but with dressed Green's function lines. 
Similarly the interactions are often dressed to become screened interactions $W$ which
now can connect vertices on different branches of the contour. Since the Green's function
$G^{(2) <}$ has the same time on the ingoing and outgoing vertex the Green's function lines
are commonly drawn closed back upon themselves to form triangles.
\begin{figure}
\includegraphics[width=0.48\textwidth]{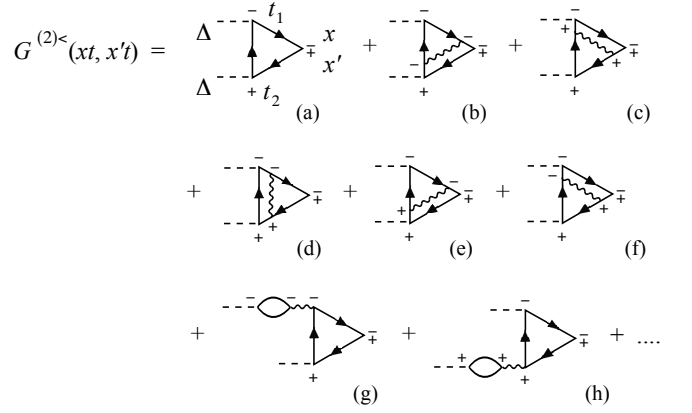} 
\caption{Skeleton expansion of $G^{(2) <}$ in $G$ and $W$ to the 
first order in $W$. }
\label{GW_tr} 
\end{figure}
In Fig.~\ref{GW_tr} we show the skeleton diagram expansion of $G^{(2) <}$ to 
lowest order in the screened interaction $W$ and the dressed Green's function $G$. 
Diagrams (a) - (c) are so-called no-loss diagrams whereas
diagrams (d) - (f) describe energy losses of the photo-electron while 
leaving the sample. Diagrams (g) and (h) describe the renormalization of the photon-field
inside the sample.  For a more in-depth discussion we refer to References [\onlinecite{Almbladh1}] 
and [\onlinecite{Almbladh3}].

\subsection{Spectral representation of the photocurrent}
Let us study the lowest order diagram in $W$ of Fig.~\ref{GW_tr}.
The structure of this diagram will also be relevant for the density functional case.
It has the explicit form
\begin{align}
&G^{(2)<} (\bx t, \bx' t)  = - \int_{-\infty}^t dt_1 dt_2 \nonumber \\
& \times
\langle \bx | \hG^{--} (t,t_1) \hD (t_1) \hG^{-+} (t_1,t_2)\hD (t_2) \hG^{++} (t_2,t) | \bx' \rangle
\label{A_diag}
\end{align}
where the minus sign originates from integration on the $+/-$ branch and 
we used the convenient notation
\be
G_{\alpha  \alpha'} (\bx t, \bx' t') = \langle \bx | \hG^{\alpha  
\alpha'} (t,t') | \bx' \rangle. \nonumber
\ee
Now the lesser Green's function $\langle \bx | \hG^< | \by \rangle$ vanishes for spatial coordinates
outside the sample as it depends only on the occupied states of the unperturbed system. We can
therefore write in our case that
\begin{align}
\hG^{--} (t,t') &= \theta (t-t') \hG^{>} (t,t') = \hG^{R} (t,t') ,\nonumber \\
\hG^{++} (t,t') &= \theta (t'-t) \hG^{>} (t,t') =- \hG^{A} (t,t'), \nonumber
\end{align}
where the retarded and advanced propagators are defined as
\begin{align}
\hG^R (t,t') &= \theta (t-t') [ \hG^> - \hG^< ] (t,t') ,\nonumber \\
\hG^A (t,t') &=- \theta (t'-t) [ \hG^> - \hG^< ] (t,t') .\nonumber
\end{align}
In terms of these propagators the expression (\ref{A_diag}) attains
the form
\begin{align}
&G^{ (2)<} (\bx t, \bx' t)  = \int_{-\infty}^t dt_1 dt_2  \nonumber \\
&\times  
\langle \bx | \hG^{R} (t,t_1) \hD (t_1) \hG^{<} (t_1,t_2) \hD (t_2) 
\hG^{A} (t_2,t) | \bx' \rangle.
\label{A_diag2}
\end{align}
This expression is valid for general time-dependent perturbations. Let us, however,
restrict ourselves to a mono-chromatic perturbation for $t > t_0$ of the form
\be
\hD (t) = \hw \, e^{-i \Omega t} + \hw^\dagger \,  e^{i \Omega t}= 
\sum_{\rho=\pm} \hw_\rho \, e^{i \rho \, \Omega t}
\nonumber
\ee
where $\Omega >0$ and where we define $\hw_{-} =\hw$ and $\hw_{+} = \hw^\dagger$.
Inserting this expression into Eq.~(\ref{A_diag2}) and assuming that $t_0$ is very far into the past
we then obtain that
\begin{align}
&G^{ (2)<} (\bx t, \bx' t)  = 
 \sum_{\rho,\eta=\pm} e^{-i (\eta + \rho) \Omega t} 
\nonumber \\
& \times
\int \frac{d\omega}{2 \pi} \langle \bx | \hG^R (\omega + \eta \Omega) \hw_\eta \hG^< (\omega) \hw_\rho \hG^A (\omega-\rho \Omega) | \bx' \rangle 
\label{G_exp}
\end{align}
where we wrote the Green's functions as Fourier transforms
\be
\hG (t,t') = \int \frac{d\omega}{2 \pi} e^{-i \omega (t-t')} \hG 
(\omega) .\nonumber 
\ee
We can now manipulate this expression in an expansion in terms of the free particle Green's functions $\hG_0^{R,A}$
of the photo-electron leaving the sample. After some manipulations which are presented in the Appendix we
find that outside the sample the lesser Green's function attains the form
\begin{align}
&G^{ (2)<} (\bx t, \bx' t)  =  \frac{\delta_{\sigma \sigma'}}{4 \pi^2 |\br | |\br'|}  \nonumber \\
& \times \int\frac{d \omega}{2\pi} e^{i q (|\br| - |\br'|)}
\langle \varphi_{q \hat{\br}} | \hw^\dagger \hG^{<} (\omega) \hw | \varphi_{q \hat{\br}'} \rangle
\label{Gless_eqn}
\end{align}
where  $q^2/2=\omega + \Omega$ is the kinetic
energy of the photo-electron and $\hat{\br} =  \br /|\br|$ is the unit vector pointing from the sample
to the detector. If we further define $\bq = q \hat{\br}$ then
the state $| \varphi_\bq \rangle $ is a quasi-particle state satisfying the equation
\be
[ \hat{h} + \hat{\Sigma}^A (\omega + \Omega) ]  | \varphi_\bq \rangle = (\omega + \Omega)  | \varphi_\bq \rangle
\label{qp_eq}
\ee
where $\hat{h}$ is the one-body part of $\hH_0$ and $\hat{\Sigma}^A$ is the advanced self-energy.
We can now use Eq.~(\ref{j_calc}) to calculate the current density which gives
\be
\mathbf{j}^{(2)} (\br t) = \frac{\hat{\br}}{4 \pi^2 |\br|^2} F(\hat{\br})
\label{Eq_curr}
\ee
where
\be
F (\hat{\br}) =  \int_{-\infty}^\mu \frac{d\omega}{2 \pi} \, q \,
\langle \varphi_{\bq} | \hw^\dagger \hat{\mathcal{A}} (\omega) \hw | \varphi_{\bq} \rangle
\label{F_func}
\ee
where we neglected terms of order $1/|\br|^3$, and used the 
fluctuation-dissipation relation $\hG^< (\omega)=i f(\w-\mu)  \hhA
(\omega)$ with $f$ the Fermi function at chemical potential $\mu$ and 
$\hhA (\omega)=i[\hG^R(\w) -\hG^A(\w)]$ the spectral function.

In the experiment one measures the flux of the current through a
space angle $d\bar{ \Omega}$ 
\be
J_{d\bar{\Omega}} = \int_{d\bar{\Omega}} \mathbf{j} \cdot d\mathbf{S} \nonumber
\ee
through a spherical surface $\mathbf{S}$ of radius $|\br|$.
If we further define $\epsilon=q^2/2=\omega +\Omega$ to be the kinetic energy of the photo-electron
as a new variable, then we can write for the current per space angle
\be
\frac{\prt J}{\prt \bar{\Omega}} (\hat{\br}) = \frac{1}{4 \pi^2 } 
\int_{-\infty}^{\mu+\Omega} \frac{d\epsilon}{2 \pi} \, \sqrt{2 \epsilon}\,
\langle \varphi_{\bq} | \hw^\dagger \hat{\mathcal{A}} 
(\epsilon-\Omega) \hw | \varphi_{\bq} \rangle.
\label{dJdO}
\ee 
Now, in an experiment also the kinetic energy of the photo-electron can be measured.
In this way the photocurrent can be split into energy contributions and we can then write
\be
\frac{\prt^2 J}{\prt \bar{ \Omega} \prt \epsilon} (\hat{\br}) =
\frac{ \sqrt{2 \epsilon} }{(2\pi)^3}  
\, \langle \varphi_{\bq} | \hw^\dagger \hat{\mathcal{A}} 
(\epsilon-\Omega) \hw | \varphi_{\bq} \rangle.
\label{dJdOde}
\ee
By measuring both the direction and energy of the photo-electron the right hand side
of this expression can be measured. 
\section{Density functional theory for photo-emission}
\subsection{Current density functional theory}

In this section we give a short overview of the basic equations of TDCDFT and its 
connection to many-body perturbation theory. A much more detailed exposition can 
be found in references [\onlinecite{tdcdft_keldysh}] and [\onlinecite{tdcdft_keldysh2}].
So far we did not specify the precise form of the perturbation.  In general its form will
be given by that of an electromagnetic field described by a time-dependent scalar potential 
and a vector potential $\bA$. 
However, we can always choose a gauge where the time-dependent fields are absorbed in
a vector potential. Static potentials, such as the potentials due to atomic nuclei, may still be
described in terms of a scalar potential absorbed in $\hH_0$.
If we do this we can write the perturbation as
\begin{align}
\hD (t) &= \int d\br \, \hat{\bj}_\pp (\br) \cdot \bA (\br,t)
 +  \frac{1}{2} \int d\br \, \hat{n} (\br)  \bA^2 (\br,t) .  \nonumber
\end{align}
We can then define a functional $\tilde{F} [\bA]$ of the vector potential by
\be
\tilde{F} [\bA] = i \ln \langle \Phi_0 | T_\gamma \, e^{-i \int_\gamma \, dz \hH (z)} | \Phi_0 \rangle  \nonumber
\ee
where $T_\gamma$ denotes contour ordering on the Keldysh contour $\gamma$ with
contour time $z$ \cite{SvL_book}.
This functional has the derivative
\be
\frac{\delta \tilde{F }}{\delta \bA (\br,z) } = \bj_{\rm p} (\br, z) + n(\br,z) \bA (\br,z) = \bj (\br,z) \nonumber
\ee
where $\bj_{\rm p}$ is the paramagnetic current and $\bj$ is the physical gauge-invariant current.
This physical current is the central object of time-dependent current-density-functional theory \cite{Ullrichbook,Vignale}.
We can make a current functional $F[\bj]$ by a Legendre transform
\be
F [\bj] = - \tilde{F} [ \bA ] +  \int_\gamma d\br dz \, \bj (\br,z ) \cdot \bA (\br ,z)  \nonumber
\ee
which has the property
\be
\frac{\delta F }{\delta \bj (\br,z)} = \bA (\br, z)  .  \nonumber
\ee
The whole derivation did not depend on the specific form of the many-body interactions in
$\hH_0$. The only thing that we assumed was that there is a one-to-one relation between the
physical current and the vector potential in our specific gauge given the initial state $| \Phi_0 \rangle$
\cite{Vignale, Tokatly}.
We could therefore have done the same derivation for a non-interacting system with initial
state $| \Phi_{0,s} \rangle$ and obtain a
current functional which we call $F_s [\bj]$. We now assume that the functionals $F [\bj]$ and
$F_s [\bj]$ have the same domain. 
We then define the exchange-correlation (xc) current functional $F_\xc$  as
\be
F_\xc [\bj] = F_s [\bj] - F [\bj] - F_\textrm{H} [\bj]
\label{xc-def}
\ee
where
\be
F_\textrm{H} [\bj] = \frac{1}{2} \int d\br d\br' \int_\gamma dz \, n(\br,z) n (\br,z) \, v(\br-\br') \nonumber
\ee 
where $v$ is the two-body interaction and
where the density $n(\br,z)$ is regarded a functional of the current through the continuity equation.
Differentiation of Eq.~(\ref{xc-def}) with respect to $\bj$ gives
\be
\bA_\xc = \bA_s - \bA - \bA_\textrm{H}  
\label{Aa}
\ee
where we defined $\bA_\xc = \delta F_\xc /\delta \bj$ and  $\bA_\textrm{H} = \delta F_\textrm{H}/\delta \bj$.
The potential $\bA_s$ is the vector potential that for a non-interacting system gives the current density $\bj$.
This system is called the Kohn-Sham system and $\bA_s$ will be called the Kohn-Sham vector potential.
If we take the initial state $|\Phi_{s,0} \rangle$ to be a Kohn-Sham ground state then the current can be
calculated by solving the Kohn-Sham single-particle equations
\be
\left[  \frac{1}{2} \big(-i \nabla + \bA_s (\br,t)\big)^2 + v_{\textrm{ext}} (\br) \right]
\phi_j  (\br,t) = i  \prt_t \phi_j (\br,t).
\nonumber
\ee
where $v_{\textrm{ext}}$ is the static external field of the unperturbed system.
If we differentiate Eq.~(\ref{Aa}) with respect to $\bj$ we obtain
\be
\frac{\delta A_{\Hxc,\mu} (\br,z)}{\delta j_\nu (\br',z')} = \frac{\delta A_{s,\mu} (\br,z)}{\delta j_\nu (\br',z')} - \frac{\delta A_\mu (\br,z)}{\delta j_\nu (\br',z')}
\label{A_j}  \nonumber
\ee
where $\bA_{\Hxc}$ is the sum of the Hartree and xc vector potentials. The indices $\mu$ and $\nu$ label the three components
of the vectors.
The quantity on the left hand side is usually called the Hartree  and xc kernel $f_{\Hxc}$ which can be split naturally  into
a Hartree part $f_\textrm{H}$ and an xc part $f_{\xc}$. The derivatives $\delta A_\mu /\delta j_\nu$ represent the inverse
of the current-current response function given by
\begin{align}
 \chi_{\mu \nu} (\br z, \br' z') &= \frac{\delta j_{\mu} (\br,z)}{\delta A_{\nu} (\br',z')} 
= \delta_{\mu \nu} n_0 (\br) \delta (\br - \br '  ) \delta (z,z')
\nonumber \\
 & -i \langle \Phi_0 | T_\gamma \left\{ \Delta \hat{j}_{{\rm 
 p},\mu,H} (\br,z) \Delta \hat{j}_{{\rm p},\nu, H} (\br' z') \right\} | \Phi_0 \rangle \nonumber
\end{align}
where the first part arises from the diamagnetic current and the second from the paramagnetic one and
we further defined  the current fluctuation operator by  $\Delta \hat{j}_{{\rm p},\mu,H} = 
\hat{j}_{{\rm p},\mu,H} - \langle \hat{j}_{{\rm p},\mu,H} \rangle$.
We have a similar response function $\chi_s = \delta j / \delta A_s$ for the Kohn-Sham system.
From Eq.~(\ref{A_j}) we see then that we can write
\be
\chi = \chi_s + \chi_s \cdot f_\Hxc \cdot \chi  \nonumber
\ee
where the dot product indicates a matrix product with respect to the indices and integration over space-time variables
on the contour. 
This is the central equation of density functional theory for linear response \cite{Ullrichbook}.
Approximations for $f_\Hxc$ can be found by expanding 
$F_\xc$ in diagrams. 
Some explicit examples of this will be given below.

\subsection{Photo-emission in current-density functional theory}
\begin{figure}
\includegraphics[width=0.48\textwidth]{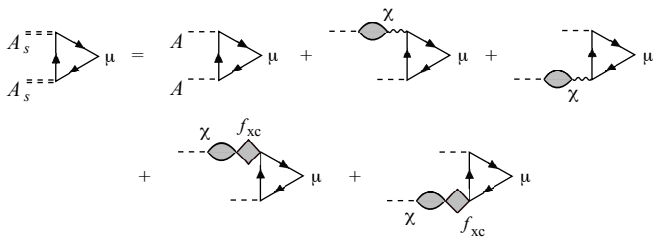} 
\caption{Diagrammatic expansion for the photocurrent in TDCDFT to linear order in 
$f_\Hxc$ and $\chi$ (see also Fig.~\ref{GW_tr}). The external vertex has a label $\mu$
corresponding to the action of a current operator and the Hartree kernel $f_{\textrm{H}}$ is indicated by 
a wiggly line.}
\label{tddft_triangle}
\end{figure}

The photocurrent within TDCDFT can be calculated as
\be
\mathbf{j}_{s}^{(k)} (\br,t  ) = \sum_{n+m=k} \langle \Psi^{(n)}_{s} (t) | 
\hat{\bj}_\pp (\br) | \Psi^{(m)}_{s} (t) \rangle 
\ee
where were have expanded the Kohn-Sham state in powers of the 
variation of the Kohn-Sham field [cfr. Eq. (\ref{currexp})]. 
The current density of TDCDFT is exactly the same as the current 
density of the real system and therefore
$\mathbf{j} (\br,t  ) =\mathbf{j}_{s} (\br,t  ) $.
Since we are measuring the photocurrent far outside the sample and 
the initial Kohn-Sham state is a Slater determinant of Kohn-Sham 
orbitals that vanish exponentially outside the sample the terms with 
$m=0$ and/or $n=0$ do not contribute. Therefore, as in section 
\ref{sec:photocurrent} where we expanded in powers of the physical vector potential,
the lowest order contribution in the Kohn-Sham field to the photocurrent is 
\be
\bj^{(2)}_{s}  (\br,t  )  = \langle \Psi^{(1)}_{s} (t) | \hat{\bj}_\pp (\br) | \Psi^{(1)}_{s} (t) \rangle.
\ee 
Since the Kohn-Sham field $\bA_{s}[\bA]$ is to lowest order linear in $\bA$ we have that $\bj^{(2)}_{s}  (\br,t  ) = \bj^{(2)} (\br,t  ) + O(\bA^{3})$.
The difference with Eq. (\ref{lopc}) is that
the initial state $| \Phi_0 \rangle$ is the Kohn-Sham initial state 
$| \Phi_{0,s} \rangle$ and that the perturbation
$\hD$ is replaced by a Kohn-Sham perturbation $\hD_s$. 
Since there are no many-body interactions in the Kohn-Sham system the diagrammatic form
of the current is simply given by the left hand side diagram in Fig.~\ref{tddft_triangle}.

\begin{figure}[h]
\includegraphics[width=0.46\textwidth]{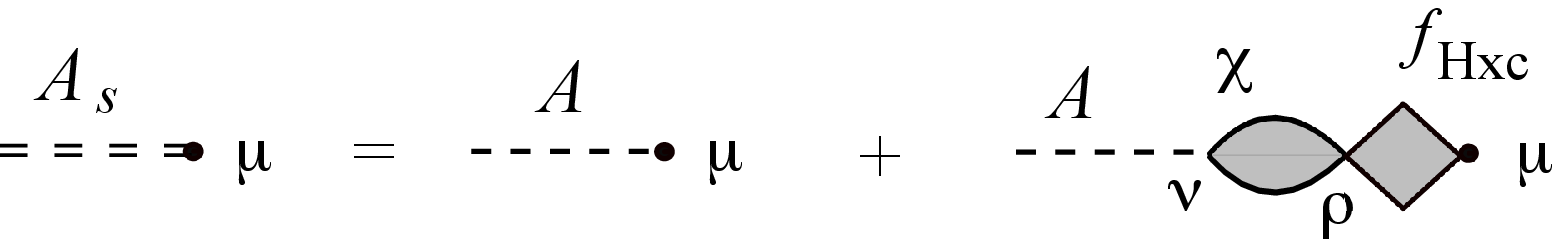} 
\caption{Diagrammatic representation of  the Kohn-Sham
field $\bA_s$ in terms of the applied filed $\bA$ (see Eq.~\eqref{A_{expansion}}).}
\label{As_eq}
\end{figure}

To write this diagram in terms of the applied field $\bA$ we need to expand the Kohn-Sham field $\bA_s$
in terms of $\bA$. To lowest order this gives
\be
A_{s,\mu} (1) = \sum_\nu \int_\gamma d2 \, K_{\mu \nu} (1,2) A_\nu  (2)
\label{A_{expansion}}
\ee
where
\begin{align}
&K_{\mu\nu} (1,2) = \frac{\delta A_{\mu,s} (1)}{\delta A_\nu (2)}   \nonumber \\
&= \delta_{\mu \nu} \delta (1,2) + \sum_\rho \int_\gamma d3 \, f_{\Hxc,\mu \rho} (1,3) \chi_{\rho \nu} (3,2)
\end{align}
and we used the short notation $j=\br_j , z_j$.
This expression can be written diagrammatically as in Fig.~\ref{As_eq}.

When we insert this diagrammatic representation in the diagrams for 
the current we obtain the graphical expansion on the right hand side 
of Fig.~\ref{tddft_triangle}, in which we only displayed terms up to linear order in  $f_\Hxc$ and $\chi$.
We note that this gives a rather different expansion than the one that we found for 
the expressions in many-body theory. In particular we see that all the
exchange-correlation contributions to the current
amount to an effective renormalization of the photon field as there are no terms connecting the
different legs of the triangle. Only the diagrams (g) and (h) in Fig.~\ref{GW_tr} have a direct correspondence
with the second and third diagram after the equal sign in Fig.~\ref{tddft_triangle} as both represent 
a renormalization due to the Hartree field. 

Let us express the Kohn-Sham current in a frequency dependent form.
If we take the external vector potential to be of the monochromatic form 
\be
\bA (\br,t) = \ba (\br) \, e^{-i \Omega t} +  \ba^* (\br) \,e^{i \Omega t}  \nonumber
\ee
then we can write
\be
\hD (t) = \hat{w} \,e^{-i \Omega t} +  \hat{w}^\dagger \, e^{i \Omega t}  \nonumber
\ee
where we neglected terms of order $\bA^2$ and defined
\be
\hat{w} = \int d\br \, \hat{\bj}_\pp (\br) \cdot \ba (\br).  \nonumber
\ee
Within linear response also the Kohn-Sham field has this form
\be
\bA_s (\br,t) = \ba_{s} (\br, \Omega) \, e^{-i \Omega t} +  \ba_s^* (\br, \Omega) \,e^{i \Omega t}  \nonumber
\ee
where
\be
a_{\mu,s} (\br,\Omega) = \sum_{\nu} \int d\br' \, K^R_{\mu \nu} (\br, 
\br'; \Omega) a_\nu (\br').  \nonumber
\ee
and where $K^R_{\mu \nu} (\Omega)$ is the Fourier transform of the 
retarded component of $K_{\mu \nu}$ evaluated at the photon frequency $\Omega$.
Then if we define 
\be
\hat{w}_s = \int d\br \, \hat{\bj} (\br) \cdot \ba_s (\br, \Omega)  \nonumber
\ee
we have that the function $F (\hat{\br})$ of (\ref{F_func}) has the following expression
in DFT
\be
F (\hat{\br}) = 
 \int_{-\infty}^\mu \frac{d\omega}{2 \pi} \, q \,
\langle \varphi_{s,\bq} | \hw_s^\dagger \hat{\mathcal{A}}_{s} 
(\omega) \hw_s | \varphi_{s,\bq} \rangle.
\label{F_func2}
\ee
Here $\hat{\mathcal{A}}_{s} (\omega)$ is the Kohn-Sham spectral function
\be
\hat{\mathcal{A}}_{s} (\omega) = 2\pi  \sum_{j}  | \phi_j \rangle  \langle \phi_j | \delta (\omega-\epsilon_j)
\label{KS_A}
\ee
where  $\epsilon_j$ and $| \phi_j \rangle$ are the Kohn-Sham one-particle energies and eigenstates
and the photo-electron orbital $| \varphi_{s,\bq} \rangle$ satisfies the equation
\be
\hat{h}_{s}  | \varphi_{s,\bq} \rangle = \frac{q^2}{2}  | \varphi_{s,\bq} \rangle  \nonumber
\ee
with incoming plane wave boundary condition.
Here $\hat{h}_{s}$ is the one-body Kohn-Sham hamiltonian of the unperturbed system.
Since the highest occupied Kohn-Sham orbital energy is equal to minus the
ionization energy, $\epsilon_N=-I$, 
(provided we choose a gauge where the potential
is zero at infinity, see Ref.~[\onlinecite{AlmbladvonBarth}]) we have $\mu=-I$ and therefore $\mu$ is the
same for the true and the Kohn-Sham system.
If we insert the expression for the spectral function operator into Eq.~(\ref{F_func2}) we find that
\be
F (\hat{\br}) =  \sum_{\epsilon_j \leq \mu} q_j \,   | \langle \phi_j   |\hw_s|  \varphi_{s,q_j \hat{\br}} \rangle |^2  \nonumber
\ee
where we defined $\epsilon_j + \Omega=q_j^2/2$. This is an exact alternative expression for $F (\hat{\br})$
of Eq.~(\ref{F_func}).
To expose its physical content we have to study explicit approximations to the xc-kernel $f_\xc$. This
will be done in the next section using diagrammatic expansions.

\subsection{Diagrammatic approximations for $f_\xc$}

\begin{figure}[t]
\includegraphics[width=0.4\textwidth]{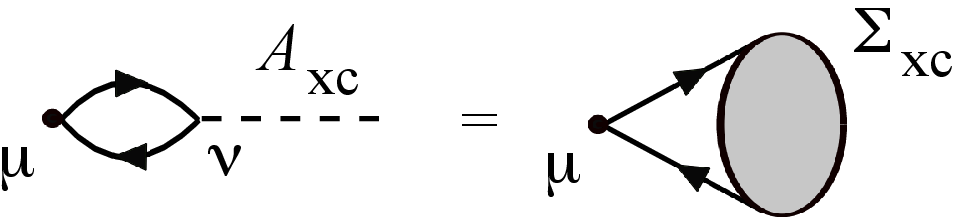} 
\caption{Diagrammatic representation of the integral equation for $\bA_\xc$ (see Eq.~\eqref{ss2}).}
\label{ss}
\end{figure}

\begin{figure}
\includegraphics[width=0.45\textwidth]{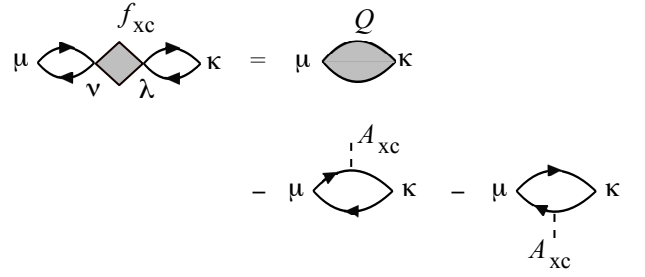} 
\caption{Diagrammatic representation for the integral equation for $f_\xc$ (see Eq.~\eqref{fxc}).}
\label{fxc_eq}
\end{figure}

\begin{figure}[h]
\includegraphics[width=0.47\textwidth]{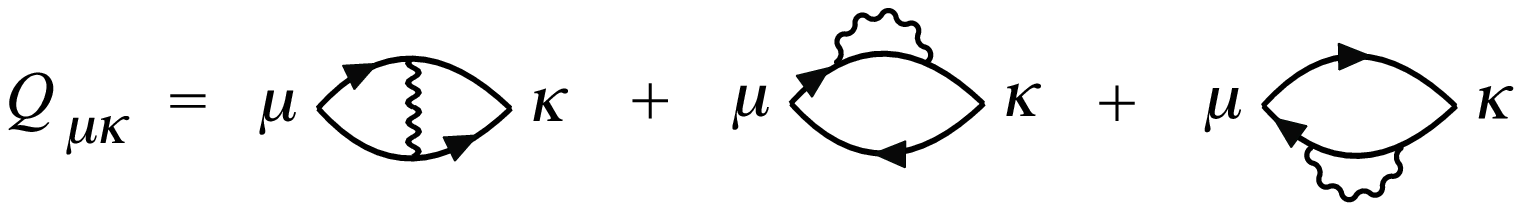} 
\caption{Exchange-only approximation to $Q_{\mu\kappa}$.}
\label{Q_eq}
\end{figure}

We will give here a brief discussion of the diagrammatic expansion
of $f_\xc$ \cite{cons_DFT,Hellgren1,Hellgren2,Hellgren3,Reining,Fukuda,Valiev,Tokatly1,Tokatly2} . 
The starting point of the discussion is the equation
\begin{align}
\frac{\delta F_{\xc}}{\delta A_{s,\mu} (1)} &= \sum_\nu \int_\gamma d2 \, \frac{\delta F_\xc}{\delta j_\nu (2)} \frac{\delta j_\nu (2)}{\delta A_{s,\mu} (1)} 
\nonumber \\
&=  \sum_\nu \int_\gamma d2 \, \chi_{s,\mu \nu} (1,2) \, A_{xc,\nu} (2)
\label{ss1}
\end{align}
where we used the symmetry in $\mu$ and $\nu$ of $\chi_s$. We now assume that $F_\xc$ is given by an
expansion in Kohn-Sham Green's functions $G_s$.\cite{note} Then the left hand side can be written as
\be
\frac{\delta F_\xc}{\delta A_{s,\mu} (1)} = 
\int_\gamma d2d3 \, \frac{\delta F_\xc}{\delta G_s (2,3)} \, G_s (2,1) \bar{j}_\mu (\br_1) G_s (1,3) 
\nonumber
\ee
where we used \cite{tdcdft_keldysh}
\be
\frac{\delta G_s (2,3)}{\delta A_{s,\mu} (1)} = G_s (2,1) \bar{j}_\mu (\br_1) G_s (1,3)  \nonumber
\ee
and defined
\be
\bar{j}_\mu (\br) = \frac{1}{2i} (\overrightarrow{\prt}_\mu -  
\overleftarrow{\prt}_\mu).
\nonumber
\ee
If we call
\be
\frac{\delta F_\xc }{\delta G_s (2,3)} = \Sigma_\xc (3,2) \nonumber
\ee
then we can write Eq.~(\ref{ss1}) as
\begin{align}
& \sum_\nu \int_\gamma d2 \, \chi_{s,\mu \nu} (1,2) \, A_{\xc,\nu} (2) = \nonumber \\
 &= \int_\gamma d2d3 \, \Sigma_\xc (3,2)  G_s (2,1) \bar{j}_\mu (\br_1) G_s (1,3)  
\label{ss2} 
\end{align}
which has the structure of a linearized Sham-Schl\"uter equation 
\cite{SS1,SS2,cons_DFT} as displayed in Fig.~\ref{ss}.
If we differentiate this equation once again with respect to $\bA_s$ we obtain an integral
equation for $f_\xc$ given by
\begin{align}
& \sum_{\nu \lambda} \int_\gamma d2d4 \, \chi_{s,\mu \nu} (1,2) f_{\xc,\nu \lambda } (2,4) \chi_{s,\lambda \kappa} (4,3)  \nonumber \\
& = Q_{\mu \kappa} (1,3) - \int_\gamma d2\, \chi_{s,\mu \nu \kappa}^{(2)} (1,2,3) \, A_{\xc,\nu} (2) 
\label{fxc}
\end{align}
where we defined
\be
 Q_{\mu \kappa} (1,3) = \frac{\delta^2 F_\xc }{\delta A_{s,\mu} (1) \delta A_{s,\kappa} (3) } 
\ee
as well as the second order Kohn-Sham response function
\be
\chi_{s,\mu \nu \kappa}^{(2)} (1,2,3) = \frac{\delta \chi_{\mu \nu} 
(1,2)}{\delta A_{s,\kappa} (3)}.  \nonumber
\ee
The corresponding integral equation for $f_\xc$ is displayed 
diagrammatically in Fig.~\ref{fxc_eq}.
The diagrammatic structure of Eq.~(\ref{fxc}) has been studied in detail in Ref.~\onlinecite{cons_DFT}
in which explicit diagrammatic expansions were derived from a Luttinger-Ward functional.
For the case of the simple exchange approximation to $\Sigma_{\xc}$, for instance, we have
\be
\Sigma_{\xc}(1,2)=-i v(1,2) G_s (1,2) 
\ee
 where $v(1,2)=\delta (z,z') v(\br_1-\br_2)$
is the bare many-body interaction. The corresponding 
diagrammatic expression for $Q_{\mu \kappa}$ is displayed in Fig.~\ref{Q_eq}.
A more advanced approximation will be discussed below.
But before we do that we first discuss the diagrammatic expansion of the
photocurrent within TDCDFT.

\subsection{Diagrams for the photocurrent}
\label{diagrams_for_photocurrent}

We have seen that we can derive approximate expression for the xc-kernel of TDCDFT
on the basis of many-body perturbation theory. A natural question to ask at this point would be
how to relate these expressions to the Feynman diagrams for the photocurrent derived directly from many-body theory,
such as the diagrams displayed in Fig.~\ref{GW_tr}. 
The situation is complicated by the fact that we do not have a direct diagrammatic expression
of $f_{\mu\nu,\xc}$ but rather one that is convoluted with two Kohn-Sham response functions 
as in Eq.~(\ref{fxc}). This is a consequence of the fact that we are 
working in the zero-temperature formalism where the  memory of initial 
correlations and initial-state dependence is lost. This allows us to 
work with time-ordered quantities that depend on the time-difference 
only
and, therefore, can be Fourier transformed. However, as it was first realized by 
Mearns and Kohn~\cite{mk.1987} and recently discussed by Hellgren
and von Barth~\cite{Hellgren2}, there are frequencies at which $\chi_s$ is 
not invertible, thus preventing a direct diagrammatic expansion of 
$f_{\xc}$ to be insert into the diagrams of Fig.~\ref{tddft_triangle}. \cite{laplacerobert}

\begin{figure}
\includegraphics[width=0.48\textwidth]{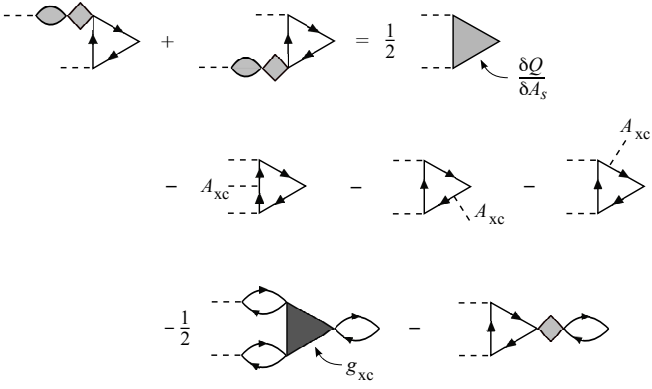} 
\caption{Diagrammatic expansion of the photocurrent obtained from differentiating the integral equation of the
xc-kernel and integrating with the external fields.  }
\label{fxc_deriv}
\end{figure}

\begin{figure}
\includegraphics[width=0.47\textwidth]{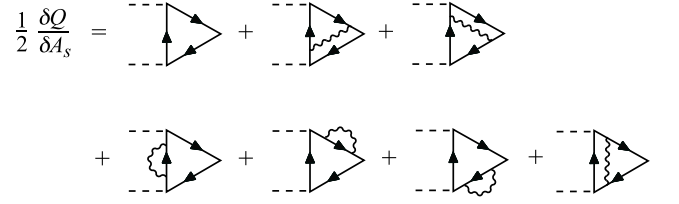} 
\caption{The exchange-only graphs contribution to the photocurrent.}
\label{fx-only}
\end{figure}

In order to generate three-point diagrams, we can differentiate Fig.~\ref{fxc_eq}
another time with respect to $\bA_s$. If we do this and collect our results we find an expression
which we display graphically in Fig.~\ref{fxc_deriv}, where after differentiation we integrated
two of the external vertices with the external field $\hD(t)$. 
In this expression we defined the higher-order
xc-kernel $g_{\xc}$ by
\be
g_{\xc, \mu \nu \tau} (1,2,3) = \frac{\delta f_{\xc,\mu \nu} 
(1,2)}{\delta j_{\tau} (3) }.
\nonumber 
\ee
The appearance of $g_{\xc}$ is not surprising given the fact that the 
photoemission problem is nonlinear in the external
field. The first filled triangle on the right hand side of  Fig.~\ref{fxc_deriv} 
represents half of the derivative $\delta Q/\delta A_s$.
For the exchange-only approximation these diagrams (integrations with 
the external field) are shown in Fig.~\ref{fx-only}.
The last two diagrams in Fig.~\ref{fxc_deriv} are exponentially small 
outside the sample as they contain the response function with a 
coordinate in the position of the detector.

Before continuing our analysis we observe that in the proximity of 
the sample the last two diagrams in Fig.~\ref{fxc_deriv} are not the only contributions to 
add to the photocurrent. In fact, the photocurrent has a first-order contribution as well
\bea
\bj^{(1)}_{s}(1) &=& \langle \Psi^{(0)}_{s} (t_1) | \hat{\bj}_\pp (\br_1) 
| \Psi^{(1)}_{s} (t_1) \rangle + c.c.\nonumber\\
&=&\int d2\; \chi_{s}(1,2)\delta \bA_{s}(2)
\label{j1}
\eea
which can be discarded only for $|\br_{1}|\to\infty$.
Let us expand this equation up to second order in the true external field~$\bA$. We have
\bea
\delta A_{s,\mu}(1) &=& \sum_{\nu}\int \frac{\delta A_{s,\mu}(1) }{\delta A_{\tau}(2)} \delta A_{\tau}(2) d2 \nonumber \\
&+& \frac{1}{2}\sum_{\tau\rho}\int \frac{\delta^2 A_{s,\mu} (1)}{\delta A_{\tau}(2) \delta A_{\rho}(3)}\delta A_{\tau}(2) \delta A_{\rho}(3) d2d3
\nonumber\\
\label{a1}
\eea
where for the second order derivative we have 
\bea
&&\frac{\delta ^2 A_{s,\mu}(1)}{\delta A_{\tau}(2) \delta A_{\rho}(3)} \nonumber\\
&=& \sum_{\zeta\eta} \int d4 d5 \frac{\delta ^2 A_{s,\mu}(1)}{\delta j_{\zeta} (4) \delta j_{\eta}(5) } 
\frac{\delta j_{\zeta} (4) } {\delta A_{\tau}(2)} \frac{\delta j_{\eta}(5) }{\delta A_{\rho}(3)}\nonumber\\
&+& \sum_{\zeta }\int d4 \frac{\delta A_{s,\mu}(1)}{\delta j_{\zeta}(4)}\frac{\delta^2 j_{\zeta}(4)}{ \delta A_{\tau}(2)\delta A_{\rho}(3)} \nonumber\\
&=& \sum_{\zeta\eta} \int d4 d5  g_{\xc,\mu\zeta\eta}(1,4,5) \chi_{\zeta\tau}(4,2) \chi_{\eta\rho}(5,3) \nonumber \\
&+&  \sum_{\zeta }\int d4 f_{{\rm Hxc},\mu\zeta}(1,4) \chi_{\zeta\mu\tau}^{(2)}(4,2,3).
\label{a2}
\eea
By inserting this back into the equation \eqref{a1} and then into Eq. 
(\ref{j1}) we obtain diagrams with the same structure as the last two 
diagrams of Fig.~\ref{fxc_deriv} but now the exact response function 
appears. Replacing the exact $\chi$ with $\chi_s$ we see that the 
diagram with $g_{\xc}$ cancels out whereas the diagram with $f_{\xc}$ is 
halved. As Eq. (\ref{a2}) contains $f_{{\rm Hxc}}=v+f_{\xc}$ we also 
get a diagram like the last diagram of Fig.~\ref{fxc_deriv} in which 
$f_{\xc}$ is replaced by the interaction $v$. In the many-body 
treatment this term arises from the expansion of $\bj^{(1)}(1)$ too.

\begin{figure}
\includegraphics[width=0.47\textwidth]{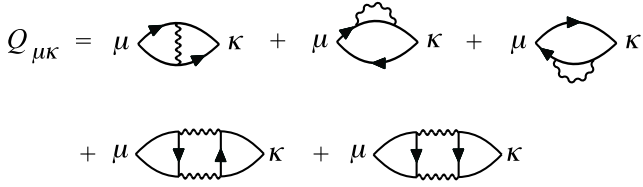} 
\caption{Diagrams for $Q_{\mu\kappa}$ in a $GW$-type approximation for $f_\xc$.
The wiggly lines denote screened interactions.}
\label{fxc_GW}
\end{figure}

\begin{figure}
\includegraphics[width=0.47\textwidth]{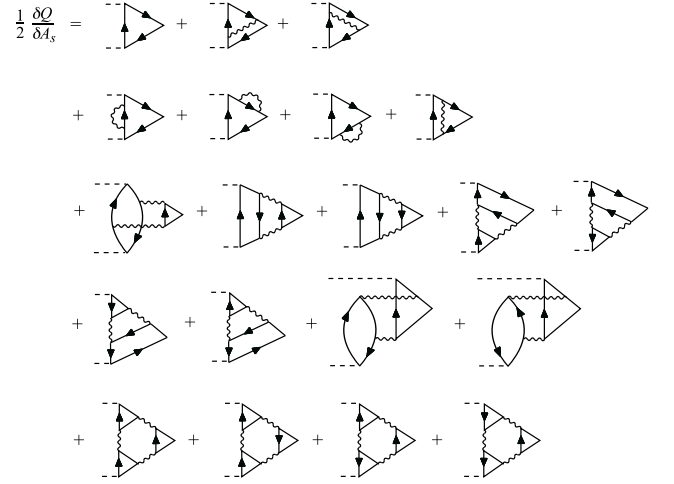} 
\caption{Diagrams for the photocurrent derived from a $GW$-type approximation for $f_\xc$.
The wiggly lines describe screened interactions.}
\label{photocurr_GW}
\end{figure}

Let us continue our analysis of the nonvanishing diagrams for the 
photocurrent outside the sample. 
From the example of Fig.~\ref{fx-only} we see that the functional 
derivative of $Q$ yields diagrams with interaction lines
connecting different legs of the triangle.  If we insert these 
diagrams back into Fig.~\ref{fxc_deriv} and then into Fig.~\ref{tddft_triangle}
we recover the expansion at the exchange-only level of the 
photocurrent (see Fig.~\ref{GW_tr} with $W\to v$)
provided we use $\chi_s$ instead of $\chi$ (this is justified since 
the expansion is first order in the interaction).  The second and 
third diagram of Fig.~\ref{tddft_triangle} are produced by a change
in the Hartree field and are also naturally included in a lowest order
many-body expansion in the bare interaction.
If we want to compare to the many-body diagrams of  Fig.~\ref{GW_tr} 
where the interaction is screened
then we also need an approximation to $f_\xc$ in terms of
$W$. To lowest order in $W$ this approximation can be derived from 
the GW self-energy
\be
\Sigma_{\xc} (1,2) = - i G_s (1,2) W(1,2) ,
\ee
where the screened interaction $W$ is the solution of 
\be
W (1,2)= v(1,2) + \int d3d4\, v(1,3) P(3,4) W(4,2) \nonumber
\ee
with polarizability $P$ given by
\be 
P(1,2) = -i G_s (1,2) G_s (2,1) . \nonumber
\ee
The corresponding equation for $Q$ is illustrated diagrammatically in Fig.~\ref{fxc_GW}.
Such an expression was studied in reference \cite{cons_DFT} from a
Luttinger-Ward functional \cite{SvL_book,KeldyshTDDFT}.  
The diagrams also include two terms which are second order in $W$ and are
important to include if one insists on having a conserving approximation \cite{cons_DFT}. 
It assures, for instance, that the $f_\xc$ satisfies the linearized zero-force theorem 
\cite{Dobson,VignaleZFT,VignaleZFT2,Mundt} which states that the exchange-correlation fields do not exert a net force
on the system.
By a differentiation of the corresponding function $Q$ and integration with the external fields we obtain the diagrams
that contribute to the photocurrent. These are displayed in 
Fig.~\ref{photocurr_GW} and have the same structure as in Fig.~\ref{GW_tr}.
We recognize all diagrams (a) - (f) of this figure. The only difference is that
we here still integrate over the two branches of the Keldysh contour.
We also note that we have some diagrams with self-energy insertions. This is because
we still expand in terms of Kohn-Sham Green's functions rather than 
the fully dressed ones.
The diagrams (g) and (h) of Fig.~\ref{GW_tr} which describe the change in
the effective Hartree field are not included in Fig.~\ref{photocurr_GW} since they are already absorbed in
the Hartree part $f_\textrm{H}$ of $f_{\Hxc}$ and are represented by the
second and third diagram after the equal sign in Fig.~\ref{tddft_triangle}.
The remaining diagrams in Fig.~\ref{photocurr_GW} describe processes that are higher order in $W$.
Such diagrams would also appear in a many-body treatment if we had expanded to
higher order in the screened interactions.  In photo-emission from metallic
systems they would, for instance, describe processes in which there are multiple
excitations of plasmons present.\cite{Ferdi,Guzzo}. 
We have seen that within TDCDFT we can make a clear connection between
the many-body expansion for the photocurrent and the Kohn-Sham expression
for it. The question that remains to be answered is whether knowledge of the
photocurrent provides us with enough information to calculate the kinetic energy
distribution of the photo-electrons. This question will be addressed in the next Section.

\subsection{Ultra-nonlocality}

\begin{figure}
\includegraphics[width=0.4\textwidth]{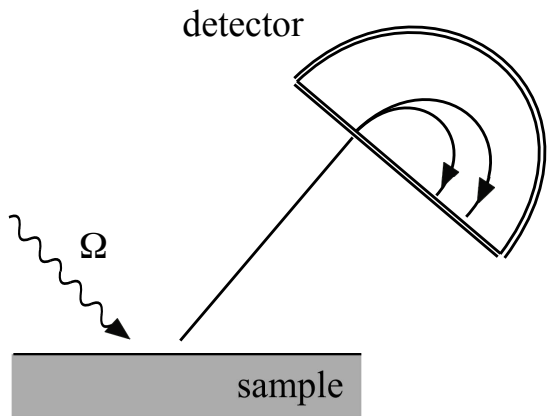} 
\caption{Deflection of two different kinetic energy components of the current by a field in the detector.}
\label{detector}
\end{figure}

In the transition from Eq.~(\ref{dJdO}) to (\ref{dJdOde}) we made a conceptual step. The total angular distribution
of the photocurrent was written as an integral over separate contributions from the electron kinetic energies $\epsilon$.
This step requires a physical interpretation since it is not justified from a mathematical point of view.
The expression can, however, be derived alternatively from Fermi's Golden Rule applied to the many-body system \cite{Almbladh2}
which amounts to a calculation of transition rates between many-body states.
We could apply the same procedure to the Kohn-Sham system but then we would calculate transitions between
Kohn-Sham Slater determinant states rather than between physical states. 
The corresponding formula would be given by removal of the integral sign in Eq.~(\ref{F_func2}) after the substitution
$\omega = \epsilon-\Omega$. This gives
\be
\frac{\prt^2 J}{\prt \bar{ \Omega} \prt \epsilon} (\hat{\br}) =
\frac{ \sqrt{2 \epsilon} }{(2\pi)^3}  
\, \langle \varphi_{s,\bq} | \hw_s^\dagger \hat{\mathcal{A}_s}
(\epsilon-\Omega) \hw_s | \varphi_{s,\bq} \rangle.
\label{dJdOdes}
\ee
It is not difficult to see that we would run into a contradiction
if we assumed that the right hand side of this equation would be identical
to the right hand side of Eq.~(\ref{dJdOde}). This becomes clearer
when we insert in Eq.~(\ref{dJdOdes}) the explicit form of the Kohn-Sham spectral function
of Eq.~(\ref{KS_A})
\be
\frac{\prt^2 J}{\prt \bar{ \Omega} \prt \epsilon} (\hat{\br}) =
\frac{ \sqrt{2 \epsilon} }{(2\pi)^2}   \sum_{\epsilon_j \leq \mu}
\, | \langle  \phi_j | \hw_s  | \varphi_{s,\bq} \rangle |^2 \, 
\delta (\epsilon - \epsilon_j - \Omega) .
\label{dJdOdes2}
\ee
If we took the example of a finite system then the spectrum on the righthand side of the equation
would only have peaks at the Kohn-Sham
energies, whereas the expression (\ref{dJdOde}) has peaks at the true removal energies of the system. We
conclude that Eq.~(\ref{dJdOdes2}) is not the same as Eq.~(\ref{dJdOde}) but that only the integrals over these functions
up to $\epsilon =\mu + \Omega$ are the same. While this is apparent 
for a finite system for an infinite system the spectral peaks
merge into a continuum and then it is not immediately obvious that 
the two expressions are different.
However, there is no reason to assume that they are equal as the interpretation based on Fermi's 
Golden Rule demonstrates. We therefore conclude that the kinetic energy distribution cannot be directly calculated
from knowledge of the current-density. This is mathematically 
clear since the momentum distribution
requires knowledge of the one-particle density matrix which is no simple functional of the current density.
However, in the experiment the kinetic energy is, in fact, measured by measuring the current at various positions
in the detector. This is done by deflecting the photo-electron 
current with an applied electric or magnetic field \cite{Damascelli}, 
as depicted graphically in Fig.~\ref{detector}. Here we display the detection of different kinetic energy components in the current.
To every position in the detector plate there is assigned a corresponding kinetic energy.
This detection process could be modeled in TDCDFT as well. 
There exists an effective Kohn-Sham field
$\bA_s$ in the region of the detector which would bend the path of the currents in exactly the same way
as the true electromagnetic field in the detector. Therefore
these kinetic energies could, in principle, also be measured in a Kohn-Sham approach. However, we realize that such a
field must have knowledge of the true many-body spectral function in the sample 
in order to split the current in exactly the right
way to produce peaks in the kinetic energy spectrum where the Kohn-Sham system has none.
This means that the exchange-correlation field in the detector far away from the sample (in fact at a macroscopic distance in a real
experiment) must depend in a nontrivial way on the many-body correlations in the sample. This is another illustration of extreme nonlocality
of the exchange-correlation field for which we can find several other instances in density-functional theory.
Other examples are the step structures in charge transfer processes in 
molecules \cite{Neepa1,Neepa2,Hellgren3}, the macroscopic exchange-correlation field
of molecular chains \cite{vanFaassen} and the lead-dependence of the exchange-correlation potential in 
quantum transport  \cite{Uimonen,KurthStefanucci,schmitteckert2013}. 

\section{Conclusions}

We derived an exact expression within TDCDFT for the photocurrent of photo-emission spectroscopy.
This expression involves an integral  over the Kohn-Sham spectral function weighted with effective
Kohn-Sham one-body interactions. Although this expression directly gives the angular dependence of the
photocurrent it does not provide us directly with the kinetic energy distribution of the photo-electrons.
This information can be obtained from TDCDFT as well, but there is a price to be paid for this. In order to
do it we need to split the photocurrent into various kinetic energy distributions using an external 
exchange-correlation field outside the sample which depends in a very nonlocal manner on the
many-body states inside the sample.  \\

From a practical point of view we may wonder whether the derived expression 
of Eq.~(\ref{dJdOdes2}) could represent a sufficiently accurate, albeit non-exact, approximation
to the kinetic energy distribution of the photo-electron spectrum. This probably depends highly on
the studied system in question. For instance, for photo-emission of metallic systems the plasmon excitations
are an important physical ingredient.  Diagrammatically these plasmonic effects  are incorporated well
in terms of Green's functions based on the GW approximation. It may well be that an xc-kernel
based on a Sham-Schl\"uter scheme at this level would give the required features in the photo-electron spectrum.
These features then would come out, not by creating extra levels in the spectral functions, but by a redistribution
of the intensities of the bare Kohn-Sham spectral function by the matrix elements involving the xc-kernel.
The most difficult case for TDCDFT is maybe provided by finite systems, such as molecules, in which non-trivial doubly or multiple excited
states \cite{Neepa3,Ruggenthaler2} may contribute important features to the spectral function. 

\acknowledgments

AMU would like to thank the Alfred Kordelin Foundation for support.
RvL would like to thank the Academy of Finland for support. GS acknowledges funding by MIUR FIRB Grant
No. RBFR12SW0J and financial support through
travel grants Psi-K2 5813
of the European Science Foundation (ESF).

\begin{appendix}
\section{Derivation of Eq. (\ref{Gless_eqn})}
We will in this Appendix give a derivation of Eq. (\ref{Gless_eqn}). We define the retarded Green's function $\hG_0^R$
for a free particle outside the sample as
\be
(i \prt_t - \hat{t} \,) \hG_0^R (t,t') = \delta (t-t')  \nonumber
\ee
where $\hat{t}$ is the kinetic energy of a free particle.
The retaded Green's function of the sample satifies
\be
(i \prt_t - \hat{h}) \hG^R (t,t') = \delta (t-t') + \int d\bar{t} \, \hat{\Sigma}^R (t,\bar{t}) \hG^R (\bar{t},t') 
\nonumber
\ee
where $\hat{h} = \hat{t} + \hat{v}$
where $\hv$ is the confining potential for the electrons in the sample
(the potential due to the atomic nuclei) and $\hat{\Sigma}^R$ is the
retarded many-body self-energy.
Then we can write the Green's function of the sample in
Dyson form as
\be
\hG^R =  \hG_0^R +  \hG_0^R (\hv + \hat{\Sigma}^R ) \hG^R  \nonumber
\ee
where integrations over internal time variables are implied.
If we now define the retarded $T$-matrix $\hat{T}^R$ by
\be
\hat{T}^R = \hv + \hat{\Sigma}^R + (\hv + \hat{\Sigma}^R) \hG_0^R \hT^R \nonumber
\ee
then we can write 
\be
\hG^R =  \hG_0^R +   \hG_0^R   \hat{T}^R  \hG_0^R .  \nonumber
\ee
If we introduce the short notations
\begin{align}
\hX^R_\eta &= (1+ \hat{T}^R \hG_0^R) (\omega+ \eta \Omega)  \nonumber \\
\hX^A_\eta &=  (1+ \hG_0^A \hT^A) (\omega+ \eta \Omega)   \nonumber
\end{align}
then we can rewrite Eq.~(\ref{G_exp}) as
\begin{align}
&G^{ (2)<} (\bx t, \bx' t)  = 
 \sum_{\rho,\eta=\pm} e^{-i (\eta + \rho) \Omega t} \int \frac{d\omega}{2 \pi} \int d\by d\by^\prime
\nonumber \\
& \times
\langle \bx | \hG_0^R (\omega + \eta \Omega)| \by \rangle
 \langle \by | \hX^R_\eta \hw_\eta \hG^< (\omega) \hw_\rho \hX^A_{-\rho} | \by' \rangle \nonumber \\
& \times  \langle \by' | \hG_0^A (\omega -\rho \Omega)| \bx' \rangle.
\label{G2_exp2}
\end{align}
Now the matrix element of $\hG_0^R$ has the explicit form
\be
\langle \bx | \hG_0^R (\nu) | \by \rangle = - \frac{\delta_{\sigma \sigma'}}{2\pi}
\left\{ \begin{array}{cc} \displaystyle \frac{e^{i \sqrt{2 \nu} \,r}}{r} &\mbox{$\nu >0$} \\ \\
 \displaystyle  \frac{e^{- \sqrt{-2 \nu} \, r}}{r} & \mbox{$\nu <0$}
\end{array} \right. \nonumber
\ee
where we defined $r=|\br-\br_1|$ with $\bx=\br, \sigma$ and $\by=\br_1, \sigma'$.
Since $\hG^< (\omega)$ has only contributions for $\omega \leq \mu$ and $\Omega >0$
we see that this matrix element only gives a contribution for $r \rightarrow \infty$
when the argument of $\hG^R$ in Eq.~(\ref{G2_exp2}) is $\omega + \Omega$.
This implies that the integral becomes
\begin{align}
&G^{ (2)<} (\bx t, \bx' t)  = \frac{1}{4 \pi^2}
  \int \frac{d\omega}{2 \pi} \int d\br d\br_2 \frac{e^{i q |\br - \br_1|}}{|\br-\br_1|}
\nonumber \\
& \times
\langle \br_1, \sigma | \hX^R_1 \hw_1 \hG^< (\omega) \hw_{-1} \hX^A_{1} | \br_2, \sigma' \rangle 
  \frac{e^{-i q |\br - \br_2|}}{|\br-\br_2|}
\label{G2_exp3}
\end{align}
where we defined $q >0$ by the relation $q^2/2 = \omega + \Omega$. If we are looking at point $\br$ far from
the sample then we can use the approximation
\be
\frac{e^{i q |\br - \br_1|}}{|\br-\br_1|} \approx \frac{e^{i q (|\br| 
- \hat{\br} \cdot \br )}}{|\br|} .  \nonumber
\ee
If we define $\bq=q \hat{\br}$ and the plane wave state $| \bq,\sigma \rangle$ with $\langle \br , \sigma | \bq , \sigma' \rangle
=\delta_{\sigma \sigma'} e^{i \bq \cdot \br}$ then we can write
\begin{align}
&G^{ (2)<} (\bx t, \bx' t)  = \frac{1}{4 \pi^2} \frac{\delta_{\sigma \sigma'}}{|\br| |\br'|}
  \int \frac{d\omega}{2 \pi}   e^{i q (|\br| - |\br'|)}
\nonumber \\
& \times
\langle q \hat{\br}, \sigma | \hX^R_1 \hw \,\hG^< (\omega) \hw^\dagger \hX^A_{1} | q \hat{\br}', \sigma \rangle 
\label{G2_exp4}
\end{align}
where we used that the Green's function must be diagonal in the spin indices.
If we then further define the state
\be
| \varphi_{q \hat{\br}} \rangle = \hX_1^A  | q \hat{\br}, \sigma \rangle = 
(1+ \hG_0^A \hT^A) (\omega+ \Omega)  | q \hat{\br}, \sigma \rangle  \nonumber
\ee
then the desired equation (\ref{Gless_eqn}) follows immediately from
Eq.~(\ref{G2_exp4}).
It remains to give a more explicit characterization of the state $| \varphi_{q \hat{\br}} \rangle$.
It satisfies the equation
\be
| \varphi_{q \hat{\br}} \rangle  = | \bq, \sigma \rangle + \frac{\hv + \hat{\Sigma} (\omega + \Omega)}{\omega + \Omega -\hat{t} -i \eta} | \varphi_{q \hat{\br}} \rangle
\label{LS_A}
\ee
which represent an advanced solution of the Lippmann-Schwinger equation with incoming plane wave
boundary conditions.
Equivalently we can write Eq.~(\ref{LS_A}) as
\be
[\hat{h} +   \hat{\Sigma}^A ( \frac{q^2}{2}) ]  | \varphi_{q \hat{\br}} \rangle   =  \frac{q^2}{2}  | \varphi_{q \hat{\br}} \rangle  \nonumber
\ee
and we see that it equivalently satisfies a quasi-particle type equation for a continuum state.
We have recovered exactly Eq.~(\ref{qp_eq}).

\end{appendix}



\begin{thebibliography}{99}
\bibitem{note}
In principle $F_\xc$  also have an explicit dependence on the
xc-field $\bA_\xc$. However, this is not the case for the approximations we consider.
\end{thebibliography}
\end{document}